\documentclass{aa}

\usepackage{txfonts} \usepackage{graphicx}
\usepackage{natbib}
\bibpunct{(}{)}{;}{a}{}{,} 

\def\dem{discrete-event model}
\def\dd{{\mathrm{d}}}
\def\zz{\phantom{0}}

\begin{document}
\title{AGN variability time scales and the discrete-event model}
\author{P. Favre\inst{1,2} \and T. J.-L. Courvoisier\inst{1,2} \and S.
  Paltani\inst{1,3}} \offprints{Thierry.Courvoisier@obs.unige.ch}
\institute{INTEGRAL Science Data Center, 16 Ch. d'Ecogia, 1290
  Versoix, Switzerland \and Geneva Observatory, 51 Ch. des
  Maillettes, 1290 Sauverny, Switzerland \and Laboratoire
  d'Astrophysique de Marseille, Traverse du Siphon, B.P. 8, 13376
  Marseille Cedex 12, France}
\date{Received ; accepted }

\abstract{We analyse the ultraviolet variability time scales 
in a sample of 15 Type 1 Active Galactic Nuclei (AGN) observed by 
{\it IUE}. Using a structure function analysis, we demonstrate the 
existence in most objects of a maximum variability 
time scale of the order of 0.02--1.00 year. We do not find any 
significant dependence of these maximum variability time scales on 
the wavelength, but we observe a weak correlation 
with the average luminosity of the objects. We also observe 
in several objects the existence of long-term variability, which 
seems decoupled from the short-term one. We interpret the 
existence of a maximum variability time scale as a possible 
evidence that the light curves of Type 1 AGN are the result of 
the superimposition of independent events. In the framework of 
the so-called discrete-event model, we study the event energy and 
event rate as a function of the object properties. We 
confront our results to predictions from existing models 
based on discrete events. We show that models based on a fixed 
event energy, like supernova explosions, can be 
ruled out. In their present form, models based on magnetic 
blobs are also unable to account for the observed relations. Stellar 
collision models, while not completely satisfactory, cannot be excluded.

}
\titlerunning{AGN variability time scales and the discrete-event model}
\maketitle

\section{Introduction}
\label{introduction}

The UV excess of the spectral energy distribution of Type 1 
Active Galactic Nuclei (AGN), the blue bump, reflects the
fact that a very large fraction of the energy is released in the
wavelength domain $\sim$ 300 to 5600 \AA\ (see e.g. \citet{krolik},
Fig.~7.10). Conventional accretion disk models are able to account
satisfactorily for the rough shape of the blue bump (but see \citet{koratkar}), 
but they fail to explain the variability properties \citep{courvoisier1}, a key to
the understanding of the AGN phenomenon.

This difficulty led several
authors (\citet{CID96}, \citet{PC97} (hereafter PC97), \citet{aretxaga}
and \citet{CID00} (hereafter CSV00); see also \citet{aretxaga2}) to 
consider a more phenomenological approach based on the
discrete-event model, which provides a simple explanation for
the variability: The variability is the result of a superimposition of
independent events occurring at random epochs at a given
rate.  The motivations for the \dem\ are twofold: Temporal analysis
allows to constrain the event properties, while its
generality leaves room for a large variety of physical events. It can
be a reasonable approximation for models like starburst
\citep{aretxaga2, aretxaga}, stellar collisions \citep{courvoisier2, 
torricelli}, or magnetic blobs above an accretion disk
\citep{haardt}.

In this paper, we use data from the {\it International Ultraviolet Explorer (IUE)}
covering about 17 years to determine the UV 
characteristic time scales in a sample of Seyfert~1 galaxies and QSOs, using a 
methodology similar to that used in \citet{collier} (hereafter CP01). While CP01 
concentrated on the measure of time scales shorter than 100 days by selecting short portions of 
the light curves in which the time sampling was denser, we use here the full available 
light curves (on average 16.5 years), highlighting a wider range of time scales, and 
extend the sample to 15 objects.  Furthermore, using 12 wavelength windows between 1300 
and 3000 \AA, we investigate for the first time the existence of a 
wavelength dependence of the variability time scale.

We interpret the variability properties of our objects in terms of \dem, and we study 
their parameters as a function of the object properties.  Our approach is similar 
to that of CSV00, although we use data from {\it IUE} gathered during almost 17 
years, while they use optical data covering about seven years.
Furthermore, the sample of CSV00 is composed of PG quasars (median $z$: 0.16) while ours is mainly composed of Seyfert~1
galaxies at a much smaller redshift (median $z$: 0.033). 
Their observations thus not only cover about three
times less time than ours, but the observation durations are further diminished in the observer's frame. Finally, as variability increases towards shorter
wavelengths \citep{kinney, paltani1, diclemente, trevese}, the study of the variability is 
more efficient in the UV than in the optical.

\section{The concept of \dem}
\label{theconceptofdem}

The formalism of the discrete-event model was mainly developed in \citet{CID96}, PC97,
\citet{aretxaga}, and CSV00. In the discrete-event model, the variability is due 
to the superimposition of independent events, occurring at random epochs, on top of a possible
constant source. In the simplest form that we use here, all events are identical. 

The total luminosity density at wavelength $\lambda$ can be expressed as:            
\begin{equation}
L_{\lambda}(t)=\sum_i e_{\lambda}(t-t_i)+C_{\lambda}\;,
\end{equation}  
where $e_{\lambda}(t-t_i)$ is the light curve of event $i$,
initiated at $t_i$, and $C_{\lambda}$ reflects the possible
contribution of a steady component. The distribution of $t_i$ is
assumed to be Poissonian.
Using $N$ for the event rate and $E_{\lambda}=\int e_{\lambda}(t-t_i)\dd t$ for 
the energy density released by the $i^{\mathrm{th}}$ event, the average luminosity reads:
\begin{equation}\label{lmean}
\overline{L_{\lambda}}=N E_{\lambda}+C_{\lambda}\;.
\end{equation}
Parameterizing the event using its duration $2\mu_{\lambda}$ and its amplitude 
at maximum $H_{\lambda}$, we have:
\begin{equation}\label{lum}
\overline{L_{\lambda}}=Nk_{\mathrm{L}} H_{\lambda}2\mu_{\lambda}+C_{\lambda}\;,
\end{equation}  
where $k_{\mathrm{L}}$ is a constant depending on the event shape.
The variance of $L_{\lambda}(t)$ was calculated by PC97 (see their equation (A5)), which reduces to
$\mathrm{Var}(L_{\lambda})\propto N$. Including  $H_{\lambda}$ and $2\mu_{\lambda}$ in Eq.~(B2) of their paper, we find:
\begin{equation}\label{var}
\mathrm{Var}(L_{\lambda})=Nk_{\mathrm{V}} H_{\lambda}^22\mu_{\lambda}\;,
\end{equation} 
where $k_{\mathrm{V}}$ is a constant depending on the event shape.
From Eqs. (\ref{lum}) and (\ref{var}), the event amplitude at wavelength $\lambda$ reads:
\begin{equation}\label{Hsix}
H_{\lambda}=\frac{k_{\mathrm{L}}}{k_{\mathrm{V}}}\frac{\mathrm{Var}(L_{\lambda})}{\overline{L_{\lambda}}-C_{\lambda}}\;.
\end{equation}

In Sect.~\ref{temporalanalysis}, we describe a method to estimate $2\mu_{\lambda}$.
As already noted by CSV00, our system will not be closed, as only three parameters can be
measured:  ($\overline{L_{\lambda}}$, $\mathrm{Var}(L_{\lambda})$, $2\mu_{\lambda}$), while our model
requires the knowledge of six parameters ($N$, $E_{\lambda}$, $2\mu_{\lambda}$, $C_{\lambda}$,
$k_{\mathrm{L}}$, and $k_{\mathrm{V}}$). Fixing the event shape determines $k_{\mathrm{L}}$ and
$k_{\mathrm{V}}$. $C_{\lambda}$ is unknown, but constrained in  the range $0\le C_{\lambda}\le \min_t
L_{\lambda}(t)$, see Sect.~\ref{thesteadycomponent}. The system can therefore be solved for these limiting cases. Under these assumptions,
we can therefore derive the energies and rates of the events from the light curves.

\section{The IUE light curves}
\label{theiuelightcurves}

\subsection{Data selection}
\label{dataselection}

We selected all the Type 1 AGN spectra available in early December 2001 
in the INES ({\it IUE} Newly Extracted Spectra) v3.0 database at VILSPA/LAEFF,
which used a new noise model and background determination \citep{rodriguez}. 
We extracted the objects monitored for several years 
for which at least 20 large aperture, small dispersion observations have been 
performed with the SWP instrument (1150--1950 \AA).  These conditions were imposed by
the temporal analysis (see Sect.~\ref{structurefunctionanalysisofthelightcurves}).
Table~\ref{Table1} gives a list of the selected objects with 
their common names and redshifts. We finally have in our sample 13 Seyfert~1 galaxies, 
one broad-line radio galaxy (BLRG), and one quasar.

\begin{table*}[tb]
\caption{ The 15 objects of the sample. 
The redshifts are taken from NED. The columns ``SWP'' and ``LWP/R'' give the number of observations in each wavelength range. 
The column ``$\Delta T$'' gives the monitoring duration in the rest frame (at 1300 \AA), while the average luminosity $\overline{L_{1300}}$ 
in the band 1300--1350 \AA\ is given in the seventh column. }
\label{Table1}
\begin{center}
\addtolength{\tabcolsep}{-2pt}
\begin{tabular}{@{}lccccccc@{}}
\hline
\hline
\rule[-0.5em]{0pt}{1.6em}
Name & Classification & $z$ & SWP & LWP/R & $\Delta T$ & $\overline{L_{1300}}$           & $E_{\rm B-V}$\\
\rule[-0.5em]{0pt}{1.6em}
     &                &     &     &       & (year)     & ($\times 10^{40}$ erg s$^{-1}$) & \\     
\hline
\object{Mrk~335} & Seyfert~1         & 0.0257 &\zz 26&\zz 28&   12.66 & 622.58 $\pm 22.78$ & 0.059\\
\object{Mrk~509} & Seyfert~1         & 0.0343 &\zz 39&\zz 32&   15.06 & 1571.8 $\pm 67.16$ & 0.060\\
\object{Mrk~926} & Seyfert~1         & 0.0475 &\zz 22&\zz 16&   14.51 & 1182.9 $\pm 137.73$& 0.053\\
\object{Mrk~1095} & Seyfert~1        & 0.0331 &\zz 35&\zz 23&   10.82 & 908.51 $\pm 29.44$ & 0.170\\
\object{NGC~3516} & Seyfert~1        & 0.0088 &\zz 71&\zz 22&   16.67 & 40.51 $\pm 2.44$   & 0.054\\
\object{NGC~3783} & Seyfert~1        & 0.0097 &\zz 95&\zz 84&   13.54 & 70.01 $\pm 2.13$   & 0.141\\
\object{NGC~4151} & Seyfert~1        & 0.0033 &   153&   137&   18.09 & 26.45 $\pm 2.17$   & 0.031\\
\object{NGC~4593} & Seyfert~1        & 0.0083 &\zz 20&\zz 15&\zz 8.31 & 12.65 $\pm 1.22$   & 0.034\\
\object{NGC~5548} & Seyfert~1        & 0.0171 &   175&   148&   16.60 & 189.70 $\pm 5.21$  & 0.024\\
\object{NGC~7469} & Seyfert~1        & 0.0163 &\zz 65&\zz 15&   17.79 & 202.78 $\pm 4.59$  & 0.079\\
\object{3C~120.0} & Seyfert~1        & 0.0330 &\zz 43&\zz 21&   15.40 & 159.96 $\pm 9.47$  & 0.160\\
\object{3C~273} & Quasar             & 0.1583 &   124&   114&   15.36 & 91782$\pm 1610.8$  & 0.027\\
\object{3C~390.3} & BLRG             & 0.0561 &\zz 99&\zz 11&   16.37 & 292.98  $\pm 16.11$& 0.071\\
\object{Fairall~9} & Seyfert~1       & 0.0461 &   139&\zz 63&   15.80 & 1981.8 $\pm 110.48$& 0.042\\
\object{ESO~141-55} & Seyfert~1      & 0.0371 &\zz 26&\zz 16&   11.33 & 1785.1 $\pm 101.40$& 0.075\\
\hline
\end{tabular}
\end{center}
\end{table*}

\begin{figure}
\includegraphics[width=9cm]{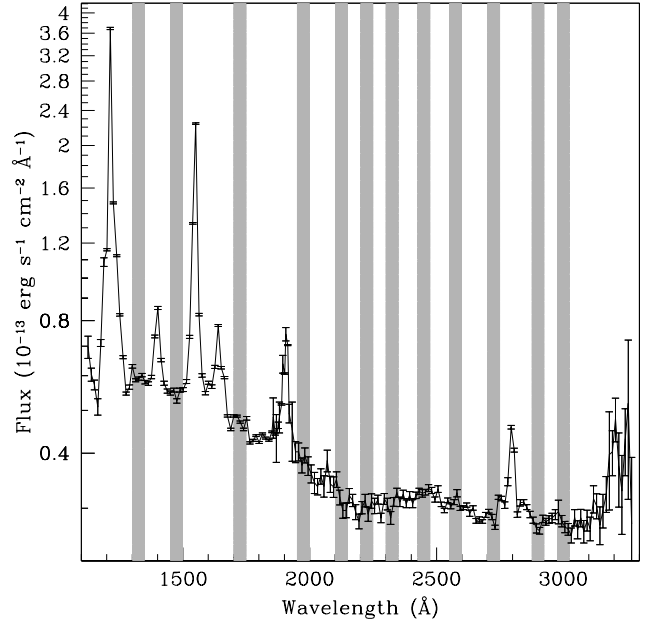} 
\caption{Average {\it IUE} spectrum of the Seyfert~1 galaxy Mrk~335. 
The position of the 12 spectral windows is indicated by the gray areas.}
\label{Fig1}
\end{figure} 

When building the light curves, FITS headers of all spectra 
were carefully checked for anomalies. We excluded spectra which were affected by
an objective technical problem stated in the FITS headers (e.g., no
significant flux detected, object out of aperture, no guiding, no
tracking). Spectra for which the pointing direction was
farther than 10\arcsec\ from the object position were also discarded. For 3C~273,
the selected spectra correspond to the list described in \citet{turler}.

\subsection{Light curves}
\label{lightcurves}

For each object, we built 12 light curves in 50 \AA\ spectral windows
starting at 1300\,\AA, 1450\,\AA, 1700\,\AA, 1950\,\AA, 2100\,\AA,
2200\,\AA, 2300\,\AA, 2425\,\AA, 2550\,\AA, 2700\,\AA, 2875\,\AA, and
2975\,\AA\ in the rest frame of the object, avoiding contamination by
strong emission lines.  Fig.~\ref{Fig1} shows the average spectrum of
Mrk~335, in which the chosen continuum spectral 
windows have been highlighted.

Two observations showing clearly spurious fluxes were removed; one in NGC~3516 
(Julian day: 2449760.78), and one in NGC~5548 (Julian day: 2448993.89).  No
correction for reddening was applied, as this should have no qualitative influence on our results (See Sect.~\ref{averageeventenergy}).

\begin{figure*}
\includegraphics[width=17cm]{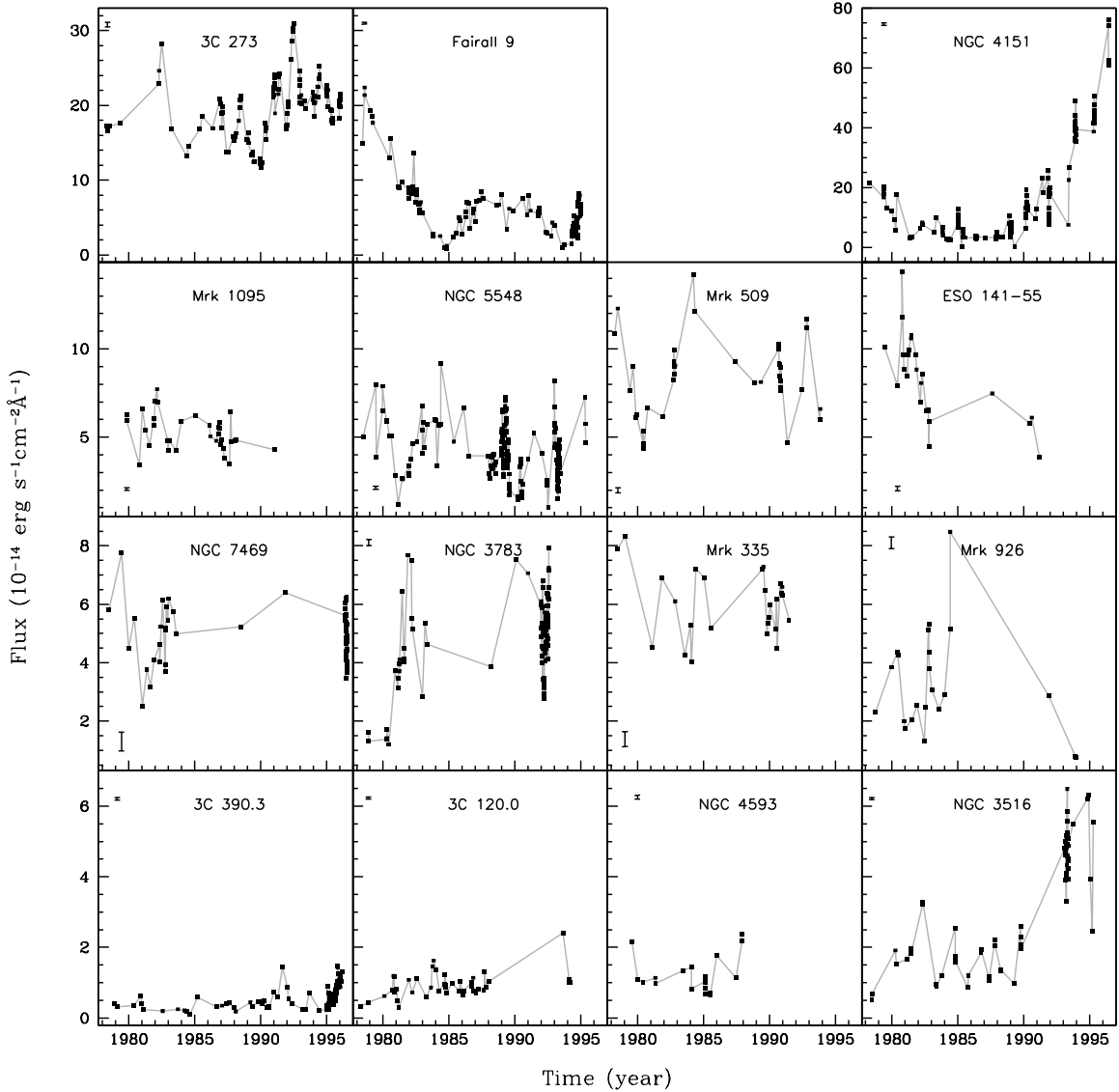} 
\caption{Light curves in the range 1300--1350 \AA\ for the 15 objects of our sample. 
Only one error bar is drawn on each panel for clarity (in the upper- or lower-left corner).}
\label{Fig2}
\end{figure*}

The light curves at 1300--1350 \AA\ for all objects in our sample are presented in Fig.~\ref{Fig2}. The 
monitoring durations in the rest frame of the objects are between 8.31 years (NGC~4593) and 18.09 years (NGC~4151), 
see Table~\ref{Table1}.  

\section{Temporal analysis}
\label{temporalanalysis}

\subsection{Structure function analysis of the light curves}
\label{structurefunctionanalysisofthelightcurves}

The first-order structure function of a light curve $x(t)$ is a
function of the time lag $\tau$, and is defined by:
\begin{equation}
{\rm SF}_{x}(\tau)=\langle (x(t+\tau)-x(t))^2\rangle\;,
\end{equation}
where $\langle y \rangle$ denotes the average of $y$ over $t$.
The structure function (hereafter SF) analysis was introduced in
astronomy by \citet{simonetti}, and is related to power
density spectrum analysis \citep{paltani4}. It measures the amount 
of variability present at a given time scale $\tau$. It has
the advantage of working in the time domain, making the method less 
sensitive to windowing and alias problems than Fourier analysis. We shall use 
here the property that, if a maximum characteristic time scale $\tau_{\rm max}$ 
is present in the light curve, the SF is constant for $\tau\ge\tau_{\rm max}$,
reaching a value equal to twice the variance of $x(t)$. Below $\tau_{\rm max}$ the SF 	
rises with a logarithmic slope of two at most. On very short time scales, 
the SF is dominated by the uncertainties on the light curve, and presents a plateau at a
value equal to twice the average squared measurement uncertainty.

We estimate the SFs of our light curves by averaging flux differences
over predefined time bins, considering only the bins containing at least six pairs. We oversample 
the SFs (i.e. the bin-to-bin interval is smaller than the bin width) in order to emphasize 
their characteristics. The bins are geometrically spaced, i.e. the bin-to-bin interval is 
constant on a Log scale.
Finally, we do not attempt to determine error bars on the 
SFs, as none of the prescriptions found in the literature
seem satisfactory.  For example, in the prescriptions of
\citet{simonetti} and CP01, the uncertainty on the SF values is proportional to $n_i ^{-1/2}$, where
$n_i$ is the number of pairs in bin $i$. These prescriptions produce underestimated error bars at
large $\tau$ (illustrated in
 Fig.~3 of CP01), because the number of pairs is increasing roughly exponentially with $\tau$ because 
of the geometric spacing of the bins, while
the total information in the light curve is finite.

\begin{figure*}
\includegraphics[width=17cm]{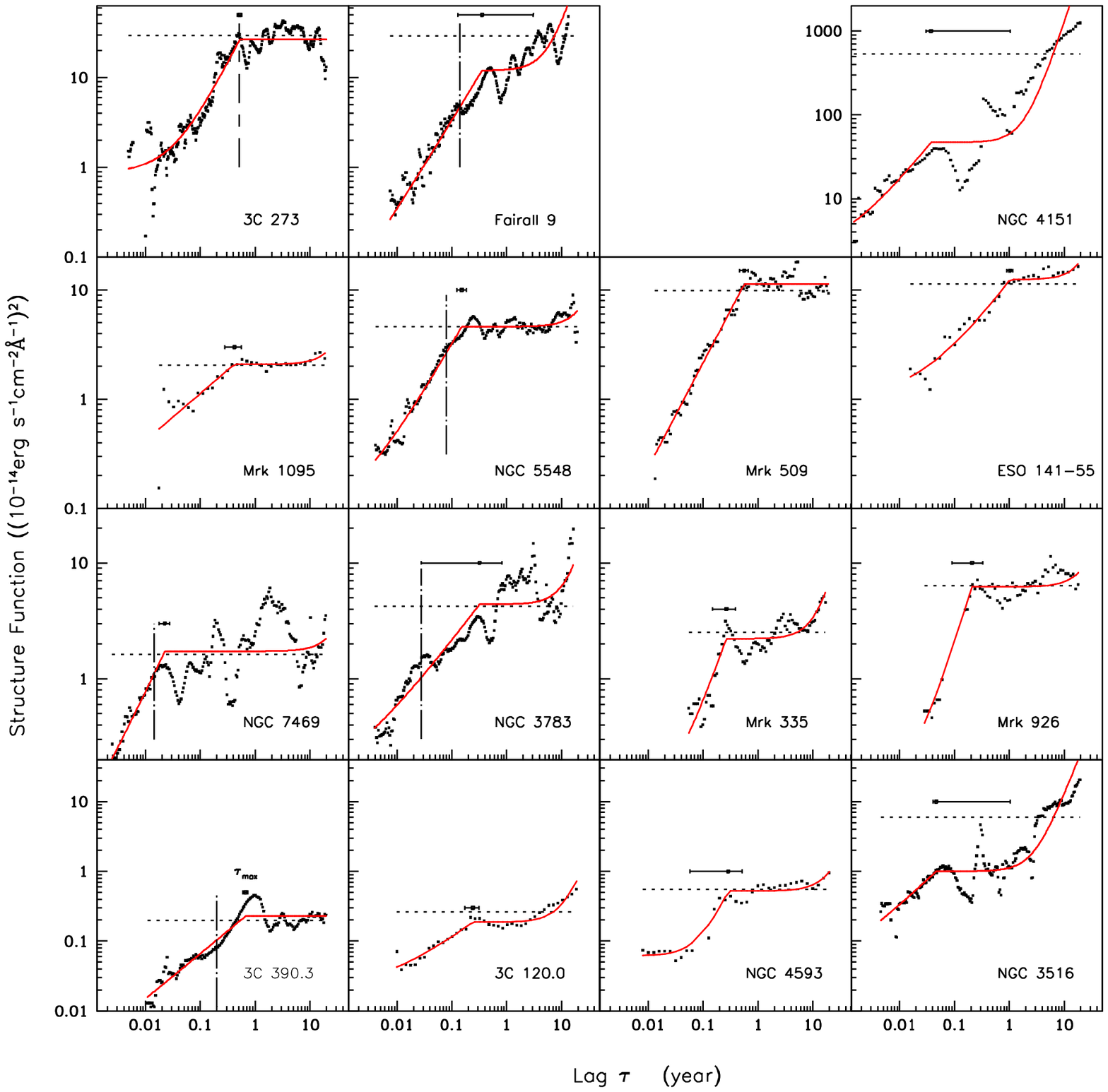} 
\caption{ Structure functions of the light curves at 1300--1350 \AA. The
horizontal dashed lines show the asymptotic value of twice the 
variance of $x(t)$.  The continuous line shows the best fit, while the value of $\tau_{\rm max}$ 
is shown with its uncertainty above the SF to ease readability. The vertical dot--long dash lines indicate the location of the characteristic 
time scales found in CP01, while the short dash--long dash line indicates the value found in \citet{paltani3}.}
\label{Fig3}
\end{figure*}

Structure functions of the 1300--1350 \AA\ light curves for the 15
objects of our sample are presented in Fig.~\ref{Fig3}. The asymptotic
values at twice the light curve variances are also shown.
Narrow structures in the SF (see e.g. Fairall~9, 3C~390.3) are very probably due
to the high inter-correlation of the SF (because the same measurements 
are used in several bins), and are probably not physical.

In Fig.~\ref{Fig3}, we observe that, for a majority of objects, the SFs show a plateau at about twice the variance of the light curves.
 However, in some cases, the SF continues to increase for $\tau$ larger than $\sim 5$ years. \citet{paltani3} presented a similar analysis on 3C~273, and 
concluded that a second component, varying on
long time scales, was present, particularly at long wavelengths. Such a component 
appears in the form of a SF rising sharply at large $\tau$.  
To cope with the possibility of the existence of a second
component, we fit the SFs using the same function as in \citet{paltani3}:
\begin{equation}
{\rm SF}_{x}(\tau)=2\epsilon^2+\left\{ \begin{array}{cc} A(\tau/\tau_{\rm max})^{\alpha}, & \tau<\tau_{\rm max}  \\
                                                            A,                               & \tau\geq\tau_{\rm max}\end{array} \right\} + B\tau^\beta\;,
\end{equation}
where $\epsilon$ is the average uncertainty on the light curves,
$\tau_{\rm max}$ the maximum variability time scale measured at the
start of the upper plateau. $A$ and $\alpha$ are the parameters of a first component of variability. 
$B$ and $\beta$ are the parameters of a second component. As we assume that this second component is slowly varying, 
we adopt $\beta=2$, the maximum value for the slope of an SF, which indicates very slow variability. 

We first fit the 1300--1350 \AA\ SFs; we shall discuss the longer wavelength SFs 
in Sect.~\ref{wavelengthdependenceoftheeventduration}. The values of $\tau_{\rm max}$
are given in Table \ref{Table2}, corrected for time dilatation. The best fits are shown on
Fig.~\ref{Fig3} by a continuous line. $\epsilon$ was fitted as a free parameter. The best fit values found for these parameters 
were comparable to the noise variance mesured in the light curves.
In four objects (NGC~3516, NGC~4151, NGC~7469,
Fairall~9), the fit does not converge unless we fix the noise
parameter $\epsilon$ and the slope $\alpha$ in the fits. We have
tested that reasonable choices of $\epsilon$ and $\alpha$ have little influence on
the value of $\tau_{\rm max}$ in those four SFs. 
We checked as well that the monitoring duration ($\Delta T$ in Table~\ref{Table1}) had no influence on the measure
of $\tau_{\rm max}$. 

The SFs of NGC~3516, and NGC~4151 show strong structures between 0.1 and 1 year,
but their behavior at $\tau>1$ is compatible with an extrapolation of their behavior at $\tau<0.1$.
Our fit interprets the strong structures as evidence of a $\tau_{\mathrm{max}}$, but we need to check
that the structures themselves do not result from the sampling. To do that, we simulated light curves using a random
walk, and projected them on the original light curve sampling. None of our simulations reproduced the
observed structures in the SFs, and we consider therefore that a maximum time scale is really present in
these two objects. The shape of the SFs forces us however to make the error bars on $\tau_{\mathrm{max}}$
extend up to 1 year in these two objects. In Fairall~9, no plateau can be seen, but there is a clear change in the slope
of the SF between $\tau=0.1$ year and $\tau=3$ years. We interpret this as the presence of a
$\tau_{\mathrm{max}}$ in this range of time scale, followed by a very strong slowly varying component.
In NGC~3783, a similar change of slope occurs, and it is impossible to locate $\tau_{\mathrm{max}}$
unequivocally. This is reflected in the error bar on $\tau_{\mathrm{max}}$ for this object. Several other objects
show the existence of a second variability component, but it does not affect the measurement of
$\tau_{\mathrm{max}}$. 

We repeated the analysis of the SFs without including a slow component of variability in the model, i.e. we fitted the data with $B=0$. The values
of $\tau_{\rm max}$ found are compared in Table~\ref{Table2}. The values found with $B=0$ are all inside the error
bars except for NGC~3516, NGC~3783 and NGC~4151. For NGC~3516 and NGC~4151, the fits do not represent the data. 
Thus for a majority of objects, the addition of a second component has no effect while it significantly improves the fits 
for NGC~3516, NGC~3783 and NGC~4151.

We conclude that with these data, we cannot decide if a second component 
is detected or not. This second variability component, while interesting per se, is outside the scope of this
paper, and shall not be discussed further.

For all the objects, we estimate the effect of the binning on $\tau_{\rm max}$ by computing and
fitting 100 SFs (at 1300--1350 \AA) for 
each object with different binnings, with corresponding bin-to-bin intervals between 0.002 year and 0.1 year.
For all objects, the distribution of the measured $\tau_{\rm max}$ is
mono-peaked, meaning that a single value of $\tau_{\rm max}$ was
always found by the algorithm. The width of this peak determines an
empirical uncertainty on $\tau_{\rm max}$ (Table~\ref{Table2}). 

We measure $\tau_{\rm max}$ from 0.022 to 0.997 year for the 15 objects
of our sample. Our time scales are of the same order of magnitude as
the one found by previous studies in the optical-UV \citep[][
CSV00]{hook, trevese94, cristiani96, paltani3, giveon}. The time scales
found by CP01 and \citet{paltani3} are indicated in Fig.~\ref{Fig3} by a vertical 
line and are in reasonable agreement with what we have found, 3C~390.3 and
NGC~3783 excepted. The discrepancy can be explained by the fact that CP01 ``detrend'' their 
SFs with a linear component (i.e. they remove a
linear fit from their light curves), arguing that the measured SF will
deviate from their theoretical shape if the light curve shows a linear
trend. Our method of including a second, slowly variable component in
the structure functions is more general than the ``detrending'', because it makes 
less strict assumptions on
the temporal properties of the slowly varying component.
It is nevertheless equivalent in the case where a linear trend is
effectively present in the data. Our method is also more
consistent in the sense that all components are handled in a similar
way. Furthermore, a linear trend would make little sense in 
objects like NGC~4151.

\begin{table}[tb]
\caption{Maximum variability time scales $\tau_{\rm max}$ of the 1300--1350 \AA\ light curves, corrected for time dilatation. The third column shows 
the values of $\tau_{\rm max}$ obtained with $B=0$, i.e. the SFs have been
fitted without a slow variability component. The last column presents the corresponding range of event durations $2\mu_{1300}$, deduced from the simulations. }
\label{Table2}
\begin{center}
  \addtolength{\tabcolsep}{-2pt}
\begin{tabular}{@{}lccc@{}}
\hline
\hline
\rule[-0.5em]{0pt}{1.6em}
Object & $\tau_{\rm max}$ & $\tau_{\rm max, B=0}$ & $2\mu_{1300}$ \\
\rule[-0.5em]{0pt}{1.6em}
      & (year) & (year) & (year) \\
\hline
Mrk~335    & 0.260$\pm 0.120$            & 0.222 & 0.238$\pm 0.169$ \\ 
Mrk~509    & 0.550$\pm 0.100$            & 0.550 & 0.489$\pm 0.381$ \\
Mrk~926    & 0.200$\pm 0.120$            & 0.227 & 0.180$^{+0.239} _{-0.120}$ \\
Mrk~1095   & 0.400$\pm 0.140$            & 0.438 & 0.242$^{+0.371} _{-0.140}$ \\
NGC~3516   & 0.046$^{+1.009} _{-0.006}$  & 4.896 & 0.050$^{+1.350} _{-0.010}$ \\
NGC~3783   & 0.319$^{+0.500} _{-0.291}$  & 1.229 & 0.300$^{+1.500} _{-0.280}$  \\
NGC~4151   & 0.037$^{+1.003} _{-0.007}$  & 2.200 & 0.040$^{+1.500} _{-0.010}$  \\
NGC~4593   & 0.284$\pm 0.230$            & 0.262 & 0.686$\pm 0.452$ \\
NGC~5548   & 0.150$\pm 0.030$            & 0.154 & 0.196$\pm 0.063$ \\
NGC~7469   & 0.022$\pm 0.005$            & 0.023 & 0.035$\pm 0.014$ \\
3C~120.0   & 0.234$\pm 0.070$            & 0.287 & 0.321$\pm 0.119$ \\
3C~273     & 0.452$\pm 0.050$            & 0.459 & 0.559$\pm 0.117$ \\
3C~390.3   & 0.631$\pm 0.070$            & 0.635 & 0.498$\pm 0.240$ \\
Fairall~9  & 0.341$^{+2.717} _{-0.227}$  & 0.760 & 0.350$^{+2.800} _{-0.250}$ \\ 
ESO~141-55 & 0.997$\pm 0.130$            & 1.014 & 1.630$\pm 0.816$ \\
\hline
\end{tabular}
\end{center}
\end{table}

\subsection{Relation with the event duration}
\label{measureoftheventduration}

An easy way to explain the existence of a maximum variability time scale is provided by the \dem.
For a Poissonian sequence of events, the SF is proportional to the SF of a single event \citep[][ CSV00]{paltani1.5,aretxaga}, 
and only has structures on time scales shorter than the event duration.
We interpret the observed SFs using discrete events for which we assume a triangular, symmetric shape.
The event shape was chosen for its simplicity \citep{paltani3}, but it
has been shown that choosing other shapes does not affect
significantly the results given by the temporal analysis (CSV00). 
Our events are described with only two parameters at wavelength $\lambda$: the event amplitude
$H_{\lambda}$ and the event duration $2\mu_{\lambda}$; $\mu_{\lambda}$ being defined as the time
needed to reach the maximum flux.
It follows that $k_{\rm L}=1/2$ and $k_{\rm V}=1/3$ in this case (see Sect.~\ref{theconceptofdem}).

\begin{figure}
\includegraphics[width=9cm]{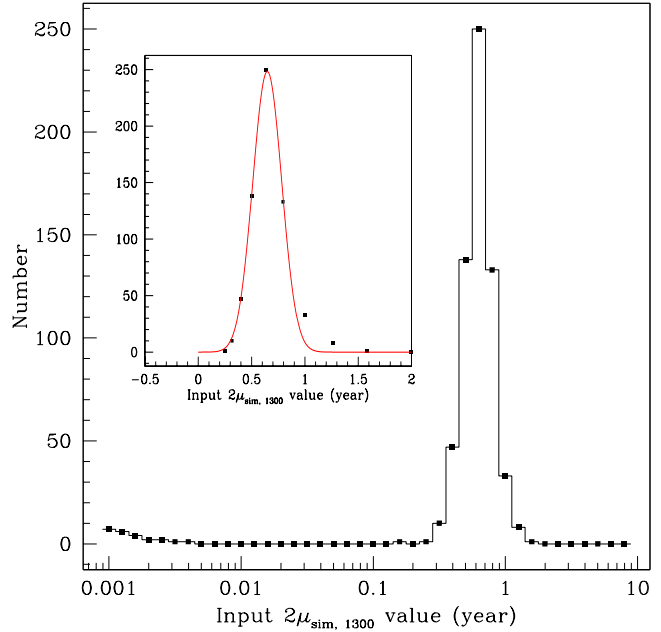} 
\caption{ Histogram of the event durations $2\mu_{\rm sim, 1300}$ input to the simulation that produce a measured $\tau_{\rm max}$ in the
range 0.4--0.5 years, for 3C~273. The inset shows the same diagram in linear scale, fitted with a Gaussian. }
\label{Fig4}
\end{figure}

While the SFs are in theory able to measure the event duration, this measure can be affected by 
the noise and sampling of the light curve in a complex and unpredictable way. To test if $\tau_{\rm max}$ measures a property of the
light curves, and not of the sampling, we produce synthetic light curves 
by simulation, and measure their $\tau_{\rm max}$.
In the simulations, we add randomly triangular events with a
given duration $2\mu_{\rm sim, 1300}$, keeping the same sampling as the
original light curve.  
A noise with an amplitude equal to the average of the instrumental noise is added to each light curve.
We take 40 test values for $2\mu_{\rm sim, 1300}$, from
$10^{-3}$ to 10 years, geometrically spaced.

\begin{figure*}
\includegraphics[width=17cm]{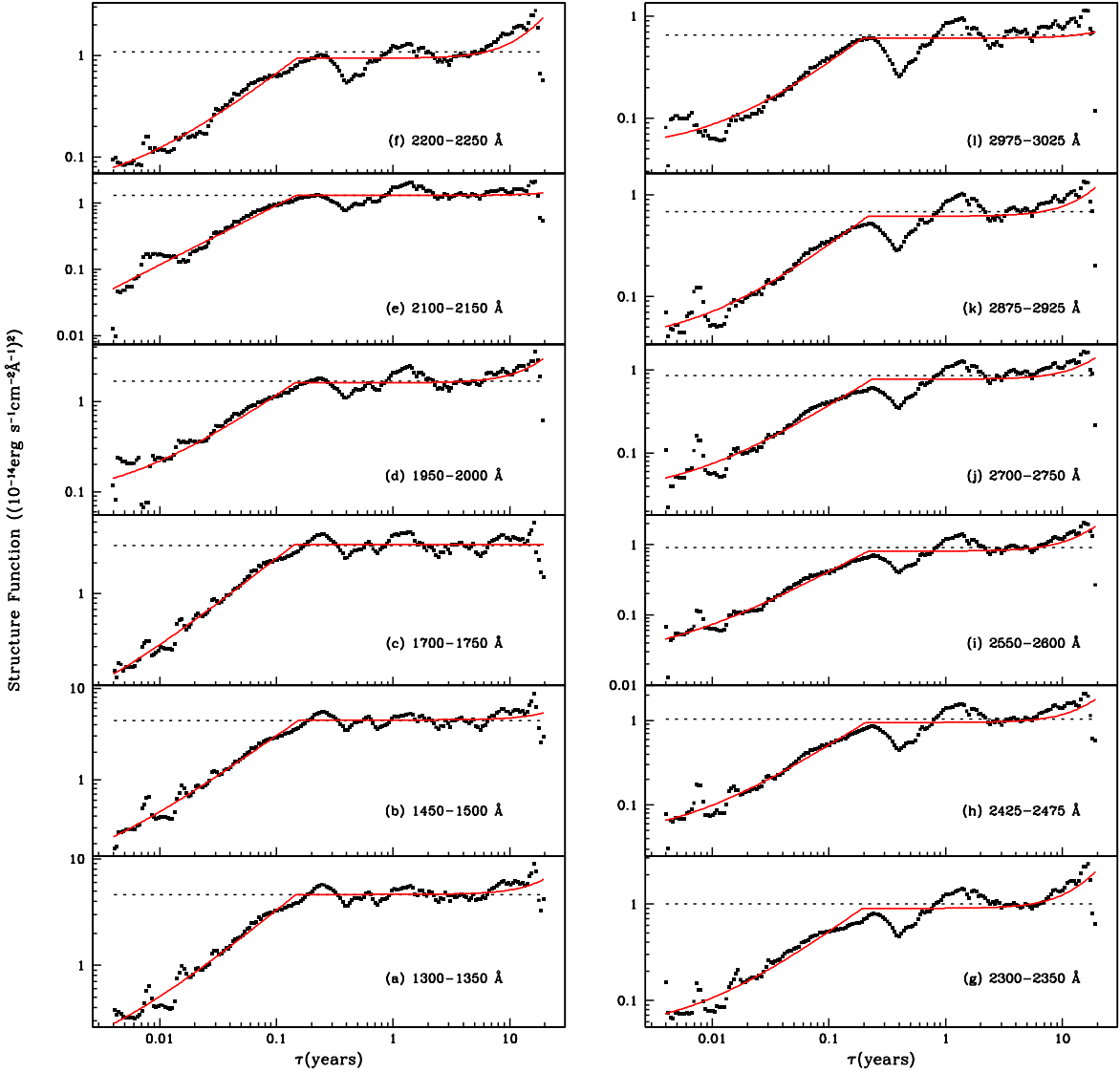} 
\caption{ SFs for the light curves 1300 to 3000 \AA, for NGC~5548. The horizontal dashed lines show the asymptotic value 
of twice the variance of the light curve. The best fits are shown by a continuous line.}
\label{Fig5}
\end{figure*}
For each object and each $2\mu_{\rm sim, 1300}$, we build 1000 light curves, compute 
their SFs and, measure $\tau_{\rm max}$ using the
method described above. The event rate $N$ is randomly chosen between 5
and 500 events per year.  
The result of the simulation is, for each object and each $2\mu_{\rm sim, 1300}$, the distribution of the
resulting $\tau_{\rm max}$. These distributions allow us to determine which input $2\mu_{\rm sim, 1300}$ can provide
the observed $\tau_{\rm max}$. For 3C~273 for example, the distribution of $2\mu_{\rm sim, 1300}$ produced
a peak around 0.56 year, as represented on Fig.~\ref{Fig4}. The fit of the peak of Fig.~\ref{Fig4} with a Gaussian
gives $2\mu_{\rm sim, 1300}=0.559\pm 0.117$ years.

We note that we never observe a $B$ parameter (see Sect.~\ref{structurefunctionanalysisofthelightcurves}) 
significantly larger than 0 in our simulations. This is  expected as we do not include a second component. 

The simulations for all objects showed a result qualitatively identical to that for 3C~273, i.e. the distributions
of $2\mu_{\rm sim, 1300}$ present a single peak. This means therefore that, for each object, $\tau_{\rm max}$
determines a unique event duration, that can be derived from the simulations. The values of $2\mu_{1300}$ are
given in Table~\ref{Table2}. We stress that our simulations are driven by the real sampling of the light curves,
and are therefore more specific than, for example, those discussed in \citet{welsh}, or CP01.

\begin{figure*}
\includegraphics[width=17cm]{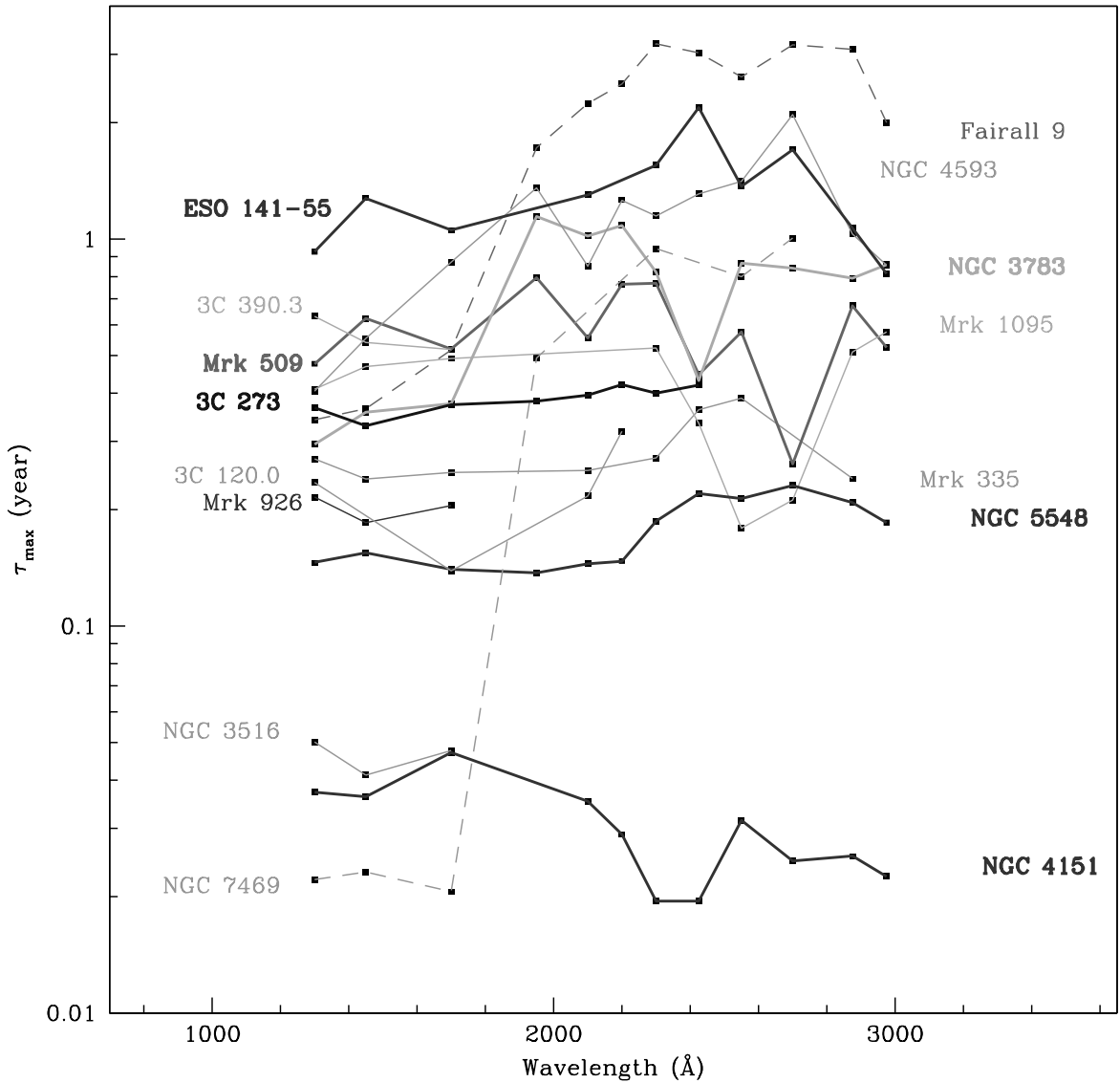} 
\caption{ $\tau_{\rm max}$ as a function of wavelength for the 15 objects. The $\tau_{\rm max}$ are corrected for time dilatation. 
The two objects which present a very strong
increase in $\tau_{\rm max}$ at long wavelengths are shown with a dashed line.}
\label{Fig6}
\end{figure*}

\subsection{Wavelength dependence of the event duration}
\label{wavelengthdependenceoftheeventduration}

For each object of the sample, we apply the method described in Sect.~\ref{structurefunctionanalysisofthelightcurves} 
to compute the event durations from the light curves 1450 to 2975 \AA.

Structure functions from 1300\,\AA\ to 3000\,\AA\ are presented in appendix~A for each object,
along with a description of the particularity of each set of SFs.  As
an example, we show the case of NGC~5548 in Fig.~\ref{Fig5}.

It is not possible to deduce a value of $\tau_{\rm max}$ for all 180
light curves. This is mainly due to the fact that, for some of the objects, the number of
observations in the LW range is too small. In addition some particular light curves are very noisy. In such cases,
the SF usually does not have the canonical shape, and the fit does not succeed. We thus reject the time scale corresponding to those
particular SFs. Each individual case is described in appendix~A. In some cases, the noise is such that, although $\tau_{\rm max}$ can be derived from the SF, its
uncertainty is very large. In such cases, one should interpret any variation in $\tau_{\rm max}$ with caution. 

Fig.~\ref{Fig6} presents the variability time scale $\tau_{\rm max}$ as
a function of the wavelength for all the objects.  We find that $\tau_{\rm max}$ is reasonably
constant over the wavelength range we use, as the small fluctuations 
can be explained by the difficulty to measure a precise
value of $\tau_{\rm max}$ on some noisy SF. For the particular case of
NGC~5548 for example, the variations are inside the uncertainties derived in the previous section. 
This result was a also found by \citet{paltani3} for 3C~273.

In two objects however, NGC~7469 and Fairall~9, the SFs present a very strong
increase in $\tau_{\rm max}$ at long wavelengths. For NGC~7469, this is due to a lack of short
term sampling of the LW light curves, which prevents the recovery of any time scale below 0.5 year. In Fairall~9, a similar lack
of short term sampling affects the determination of $\tau_{\rm max}$. However, the values of $\tau_{\rm max}$ in the LW range
are within the uncertainties on $\tau_{\rm max}$ determined at 1300 \AA.

\section{Interpretation in terms of \dem}
\label{interpretationintermsofdem}

\subsection{$2\mu_{1300}$ as a function of the luminosity}
\label{2muasafunctionoftheluminosity}

The relationship between the event duration and the luminosity
has some important consequences for the \dem\ that we will
discuss below (Sect.~\ref{constraintsonthevariabilityluminosityrelation}).
We note, however, that only a weak dependence between the event duration and the 
average luminosity of the object is possible as the values of the former cover
less than two orders of magnitude while the latter covers four orders of
magnitude. We thus measure a physical time scale which has at most a small dependence on the
luminosity of the objects.

\begin{figure}
\includegraphics[width=9cm]{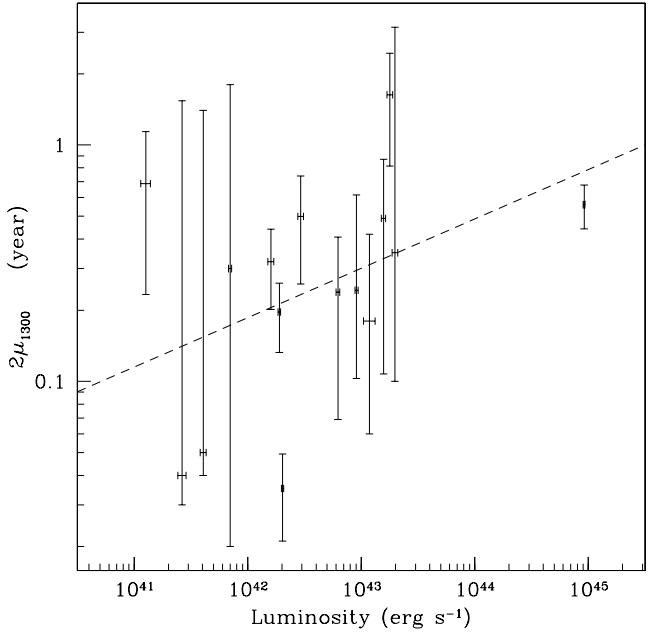} 
\caption{ Variability time scales $2\mu_{1300}$ as a function of the luminosity of the objects in the range 1300--1350 \AA. The BCES regression is shown with the dashed line.}
\label{Fig7}
\end{figure}

The event duration $2\mu_{1300}$ as a function of the
average luminosity $\overline{L_{1300}}$ of the object at 1300--1350 \AA\ is shown in
Fig.~\ref{Fig7}. We use a simple cosmology with $H_0=60$ km s$^{-1}$ Mpc$^{-1}$, and $q_0=0.5$ 
throughout.

Using Spearman's correlation coefficient, we find a correlation between the event duration and the 
luminosity ($s=0.38$), marginally significant at the 16\% level (Null hypothesis). The dependence of $2\mu_{1300}$ on $\overline{L_{1300}}$
can be expressed as $2\mu_{1300}\propto\overline{L_{1300}}^{\delta}$, where the index $\delta=0.21\pm 0.11$
has been determined using the BCES linear regressions \citep{akritas96}.

\subsection{The steady component $C_{\lambda}$}
\label{thesteadycomponent}

The steady component $C_{\lambda}$ (Sect.~\ref{theconceptofdem}) 
can have various physical origins.  For example, it can be associated to 
the non-flaring part of the accretion disk, or to the host-galaxy stellar contribution 
(\citet{CID96}, CSV00). We shall however continue our 
discussion in a model-independent way.
We can constrain $C_{\lambda}$ for a particular light curve by imposing that it does not exceed
the minimum observed luminosity $L_\lambda^{\mathrm{min}}$. On the other hand, $C_{\lambda}=0$ is an obvious lower limit (although \citet{palwa}
argued that $C_{\lambda} > 0$, at least for $\lambda>2000$ \AA). In the following, we shall use these
two constraints as limiting cases.

\subsection{Spectral shape of the event amplitude, event energy and event rate}
\label{averageeventenergy}

\begin{figure}
\includegraphics[width=9cm]{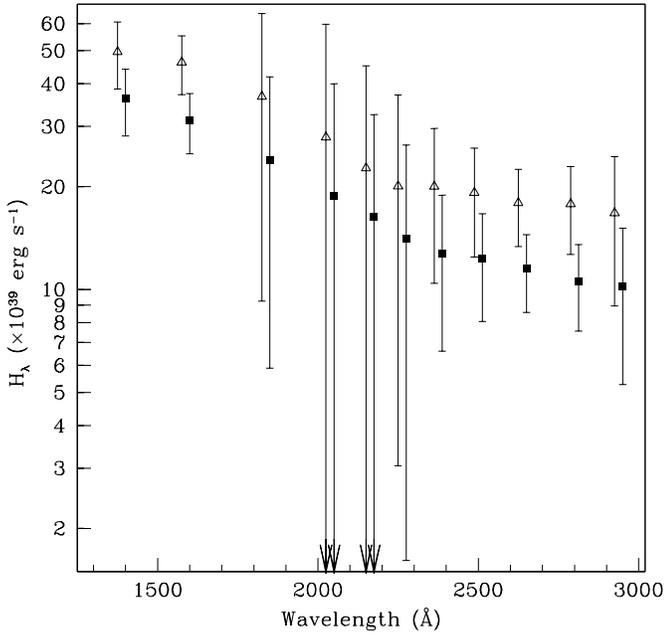}
\caption{ Event amplitude spectrum $H_{\lambda}$ as a function of wavelength for NGC~5548, with $C_{\lambda}=0$ (squares), and 
$C_{\lambda}=L_\lambda^{\mathrm{min}}$ (open triangles). 
The lower curve has been slightly shifted to the right to ease readability.} 
\label{Fig10}
\end{figure}

For each object, we compute the event amplitude $H_{\lambda}$ from Eq.~(\ref{Hsix}) 
using both $C_{\lambda}=0$ and $C_{\lambda}=L_\lambda^{\mathrm{min}}$. Fig.~\ref{Fig10} shows the 
spectral shape of $H_{\lambda}$ for NGC~5548.
We integrate $E_{\lambda}=2\mu_{\lambda}k_{\rm L}H_{\lambda}$ interpolated over the wavelength range 1300--3000 \AA\ 
to obtain the energy $E$ released in one event, assuming isotropic emission:
\begin{equation}\label{inte}
E=\int_{1300~\AA} ^{3000~\AA}E_{\lambda}\dd\lambda\;.
\end{equation}	
We use the event duration at 1300--1350 \AA, as it can be considered constant over the wavelength range considered. The
event rates $N$ are derived from Eq.~(\ref{lum}).

\begin{figure}
\includegraphics[bb=20 150 570 500,clip,width=9cm]{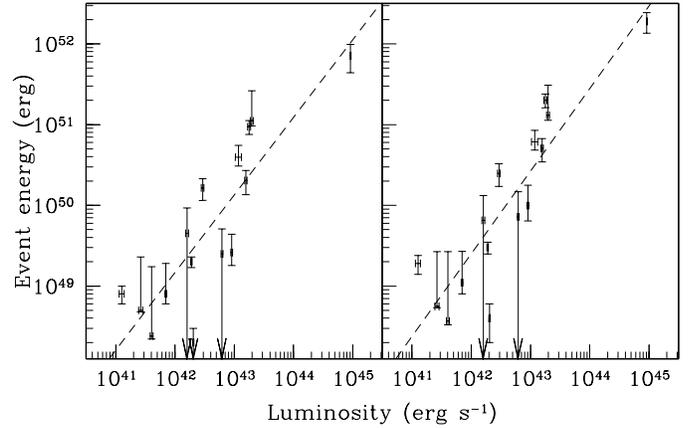}
\caption{ Event energy as a function of the object luminosity in the hypothesis $C_{\lambda}=0$ (left panel), and using the upper limit of $C_{\lambda}$
(right panel). The BCES regression is shown with the dashed line.}
\label{Fig11}
\end{figure}

Table~\ref{Table4} gives the event energies and rates found with the $C_{\lambda}=0$, and 
$C_{\lambda}=L_\lambda^{\mathrm{min}}$ assumptions. Event energies are 
found in the range $10^{48}-10^{52}$ erg, and maximum event rates in the range 
9--1133 event year$^{-1}$ ($C_{\lambda}=0$) while minimum event rates are found in the 
range 2--270 event year$^{-1}$ (upper limit of $C_{\lambda}$).
Fig.~\ref{Fig11} shows the event energy $E$ as a function of the 1300--1350 \AA\ luminosity, using the upper
and lower limits on $C_{\lambda}$.

\begin{table*}[tb]
\caption{ Event rate $N$ and energy $E$ for each object, with the $C_{\lambda}=0$, and
        $C_{\lambda}=L_\lambda^{\mathrm{min}}$ assumptions.}
\label{Table4}
\begin{center}
\addtolength{\tabcolsep}{-2pt}
\begin{tabular}{@{}lcccc@{}}
\hline
\hline
\rule[-0.5em]{0pt}{1.6em}
Name & $N$ & $E$ & $N$ & $E$ \\
\rule[-0.5em]{0pt}{1.6em}
& (year$^{-1}$) & ($\times 10^{50}$ erg) & (year$^{-1}$) & ($\times 10^{50}$ erg) \\       
 &\multicolumn{2}{c}{$C_{\lambda}=0$}&\multicolumn{2}{c}{$C_{\lambda}=L_\lambda^{\mathrm{min}}$}\\
\hline
Mrk~335      &  160.76 $\pm$ 123.78   &  0.25 $\pm$  0.26  & 17.82 $\pm$ 41.22  &  0.72 $\pm$  0.76 \\
Mrk~509      &   39.28 $\pm$ 31.89    &  2.03 $\pm$  0.67  &  9.09 $\pm$ 15.34  &  5.08 $\pm$  1.64\\
Mrk~926      &   24.84 $\pm$ 17.20    &  3.94 $\pm$  0.87  & 14.66 $\pm$ 13.22  &  6.12 $\pm$   1.30  \\
Mrk~1095     &  149.32 $\pm$ 90.68    &  0.26 $\pm$  0.08  & 18.45 $\pm$ 31.88  &  0.99 $\pm$ 0.35 \\
NGC~3516     &  104.90 $\pm$  23.42   &  0.024 $\pm$ 0.002 & 75.44 $\pm$  19.86 &  0.037 $\pm$  0.004 \\
NGC~3783     &   50.11 $\pm$  48.48   &  0.08 $\pm$  0.02  & 28.21 $\pm$  36.37 &  0.11 $\pm$  0.03\\
NGC~4151     &   32.12 $\pm$  8.62    &  0.05 $\pm$  0.002 & 31.16 $\pm$ 8.49  &  0.056 $\pm$ 0.002 \\
NGC~4593     &   10.32 $\pm$ 7.09     &  0.08 $\pm$  0.02  &  2.32 $\pm$ 3.36 &  0.19 $\pm$  0.05\\
NGC~5548     &   51.59 $\pm$ 18.49    &  0.20 $\pm$  0.03  & 29.52 $\pm$ 13.99 &  0.30 $\pm$  0.05\\
NGC~7469     &    1133.8 $\pm$ 543.73 &  0.02 $\pm$  0.01  &269.88 $\pm$ 265.39&   0.04 $\pm$  0.02\\
3C~120.0     &   27.53 $\pm$ 12.42    &  0.45 $\pm$  0.48  & 13.22 $\pm$  8.61&   0.65 $\pm$   0.67\\
3C~273       &   62.44 $\pm$  14.51   &  71.20$\pm$  27.34 & 10.24 $\pm$  5.88&   189.62 $\pm$ 53.83\\
3C~390.3     &    8.94 $\pm$  4.82    &  1.64 $\pm$  0.49  &  5.86 $\pm$  3.90&   2.49 $\pm$  0.78\\
Fairall~9    &    8.82 $\pm$  6.44    &  11.11 $\pm$ 1.49  &  6.50 $\pm$ 5.53 &  13.03 $\pm$ 1.74 \\
ESO~141-55   &    9.75 $\pm$   5.11   &  9.38 $\pm$  1.85  &  2.69 $\pm$  2.69&   20.03 $\pm$  3.86 \\
\hline
\end{tabular}
\end{center}
\end{table*}

\begin{figure}
\includegraphics[bb=20 150 570 500,clip,width=9cm]{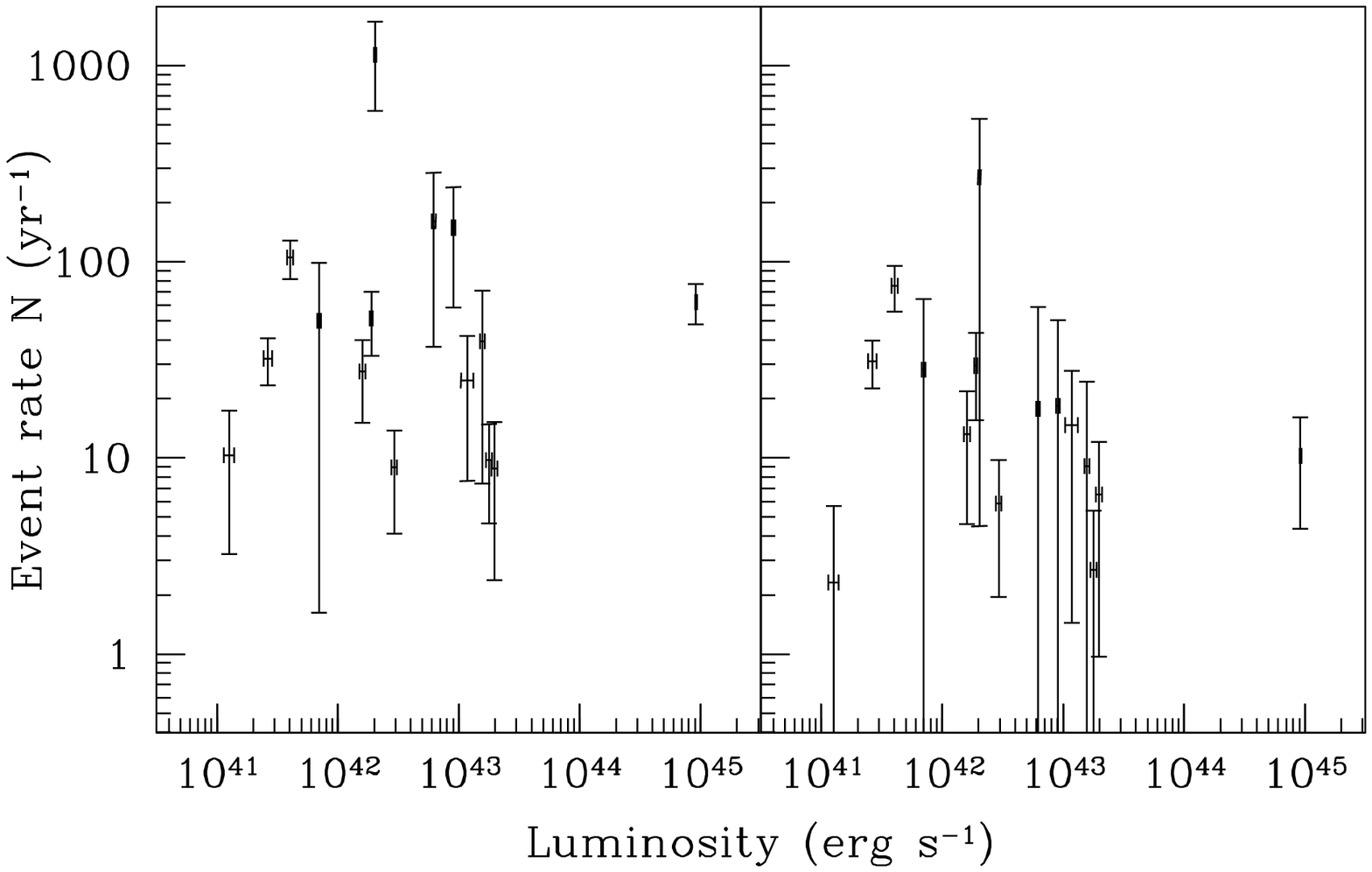}
\caption{ Event rate as a function of the object luminosity in the hypothesis $C_{\lambda}=0$ (left panel), and using the upper limit of $C_{\lambda}$ (right panel).}
\label{Fig13}
\end{figure}

With $C_{\lambda}=0$, $E$ and $\overline{L_{1300}}$ are clearly correlated (Spearman correlation coefficient: $s=0.85$), with a Null hypothesis 
probability of less than 0.1\%. A linear regression gives $E\propto\overline{L_{1300}}^{\gamma}$, with 
$\gamma=0.96\pm 0.09$ using the BCES method \citep{akritas96} (Fig.~\ref{Fig11}, left panel). 

When using $C_{\lambda}=L_\lambda^{\mathrm{min}}$, the correlation is preserved (Spearman: $s=0.89$), with a Null hypothesis probability of less
 than 0.1\% and an index $\gamma=1.02\pm 0.11$ (Fig.~\ref{Fig11}, right panel).
We note that $E$ and $\overline{L_{1300}}$ are lower limits
since no corrections for reddening were applied. The correlation should be preserved by applying the corrections, as both $E$ and $\overline{L_{1300}}$ would be scaled 
by the same factor. We tested this hypothesis by correcting the light curves for reddening and recomputing the relation $E=f(\overline{L_{1300}})$. The correlation is 
preserved (Spearman correlation coefficient: $s=0.83$) with an index $\gamma=0.93\pm 0.10$ for the case $C_{\lambda}=0$ as well as for the case 
$C_{\lambda}=L_\lambda^{\mathrm{min}}$ (Spearman correlation coefficient: $s=0.84$) with an index $\gamma=0.97\pm 0.11$. These values are consistent with
the non-dereddened values. 
 
The event rates as a function of the object luminosity are given in Fig.~\ref{Fig13}, for both the upper and lower limits on $C_{\lambda}$.
A non-significant anticorrelation is found in the $C_{\lambda}=0$ case (Spearman's $s=-0.12$, Null hypothesis probability of 66.64\%),
 while a marginally significant anticorrelation is found for $C_{\lambda}=L_\lambda^{\mathrm{min}}$ (Spearman's $s=-0.39$, Null hypothesis probability of 15\%). 
Using the data corrected for reddening, one finds as well a non-significant anticorrelation (Spearman's $s=-0.02$, Null hypothesis probability of non-correlation 
of 93.96\%) in the case $C_{\lambda}=0$, and a non-significant anticorrelation for $C_{\lambda}=L_\lambda^{\mathrm{min}}$ (Spearman's $s=-0.17$, Null hypothesis 
probability of 54.12\%).  

\subsection{Constraints on the variability-luminosity relation}
\label{constraintsonthevariabilityluminosityrelation}

\citet{paltani1} showed that the variability of a similar sample was anticorrelated
to the object luminosity. PC97 confirmed this result in the rest-frame of the objects and 
found $\sigma_{1250} ^{\rm rest}\sim\overline{L_{1250}} ^{\eta}$, with $\eta=-0.08\pm 0.16$.

We showed in Sect.~\ref{2muasafunctionoftheluminosity} that the event 
duration $2\mu_{1300}$ is a shallow function of the luminosity 
and in Sect.~\ref{averageeventenergy} that the measured event energy 
$E\propto\overline{L_{1300}}^{\gamma}$, with $\gamma\simeq 1$.
These two results, expectedly, lead to values of $\eta$ in agreement 
with the measure of PC97. We thus established that the event parameter 
which drives the $\sigma(L)$ dependence is the event energy, and not its duration, nor its rate.

PC97 showed that in this case one should expect a variability-luminosity relation in the form: 
\begin{equation}
\sigma(L)\propto\overline{L_{1300}}^{\frac{\gamma-\delta}{2}-\frac{1}{2}}\;.
\end{equation}
We present in Table~\ref{Table6} the different values of the slope $\eta$ of the 
variability-luminosity relationship which are 
consistent with the value of PC97.

\begin{table}[tb]
\caption{ Slope $\eta$ of the variability-luminosity relation for
the two $C_{\lambda}$ assumptions (see text).}
\label{Table6}
\begin{center}
\addtolength{\tabcolsep}{-2pt}
\begin{tabular}{@{}lccc@{}}
\hline
\hline
\rule[-0.5em]{0pt}{1.6em}
& $\gamma$ & $\delta$ & $\eta$\\
\hline
\rule[-0.5em]{0pt}{1.6em}
$C_{\lambda}=0$                        & $0.96\pm 0.09$ & $0.21\pm 0.11$ & $-0.13\pm 0.06$\\
$C_{\lambda}=L_\lambda^{\mathrm{min}}$ & $1.02\pm 0.11$ & $0.21\pm 0.11$ & $-0.10\pm 0.05$\\
\hline
\end{tabular}
\end{center}
\end{table}

\section{Discussion}
\label{discussion}
\subsection{Event duration and black-hole physical time scales}
\label{eventduration}

\begin{figure}
\includegraphics[width=9cm]{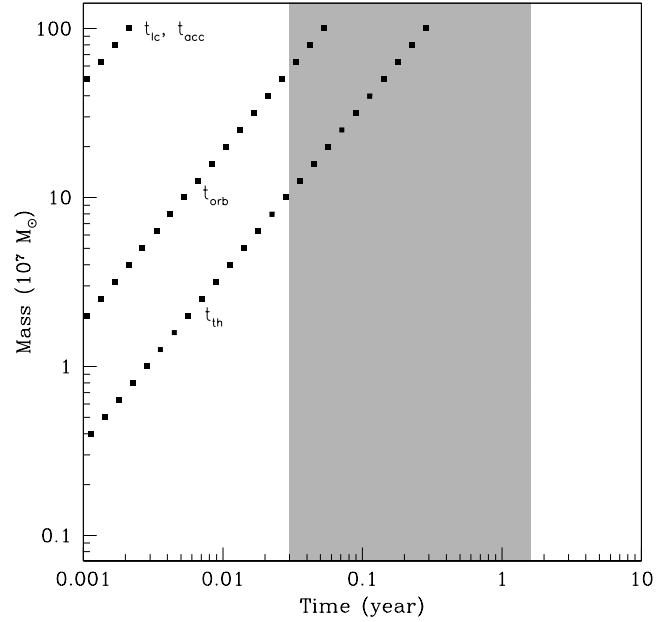}
\caption{ Time scales vs black hole mass. The gray area shows the range of timescales found in this study.} 
\label{Fig15}
\end{figure}

We have shown that a characteristic variability time scale exists, which
can be measured in the light curves. It can be associated with the
event duration in a model-independent way. We have obtained 
event durations in the range 0.03 to 1.6 years, which may possibly be related
to the four physical time scales associated to black holes. They all depend on 
the black hole mass $M_{\rm BH}$ and Schwarzschild radius $R_{\rm S}$. 
We review them below, from the fastest to the slowest, 
following \citet{edelsonnandra} and \citet{manmoto}.
\begin{enumerate}
\item The light crossing time 
is given by $t_{\rm lc}=3.01\times 10^{-5}M_7(R/10R_{\rm S})$ year;
\item the ADAF accretion time scale (comparable to free-fall velocity) is given by $t_{\rm acc}\geq 4.38\times 10^{-3}M_7(R/100R_{\rm S})^{3/2}$ year;
\item the gas orbital time scale is given by $t_{\rm orb}=9.03\times 10^{-4}M_7(R/10R_{\rm S})^{3/2}$ year;
\item the accretion disk thermal time scale $t_{\rm th}=1.45\times 10^{-2}(0.01/\alpha_{\rm visc})M_7(R/10R_{\rm S})^{3/2}$ year,
\end{enumerate}
where $R$ is the emission distance from the center of mass, $M_7=M/10^7 M_{\odot}$, and the Schwarzschild radius is defined $R_{\rm S}=2GM/c^2$.
$t_{\rm th}$ and $t_{\rm orb}$ can produce the range of time scales observed here for reasonable black hole masses 
(see Fig.~\ref{Fig15}). However, all these time scales depend linearly on the mass of the object, 
hence on the luminosity. The lack of strong dependence of the event duration on the object luminosity allows us 
to exclude all these mechanisms as likely candidates for the origin of the variability.

\subsection{Physical nature of the events}
\label{physicalnatureoftheevents}

\subsubsection{Supernovae}
\label{supernovae}

In the starburst model \citep{aretxaga}, the variability of AGN is produced by 
supernovae (SNe) explosions and compact supernovae remnants (CSNRs). The SNe generate 
the CSNR in the interaction of their ejecta with the stellar wind from 
the progenitor. \citet{terlevich92} showed that the properties of CSNR match 
the properties of the broad-line region of AGN.

\citet{aretxaga93,aretxaga2} modeled the B band 
variability of the Seyfert galaxies NGC~4151 and NGC~5548
 with this model. For NGC~4151, an event rate of 0.2--0.3 events year$^{-1}$ was found. However, 
typical predictions of the model are more of the order of 3--200 events year$^{-1}$ \citep{aretxaga}, consistent 
with what we found. The event energy has to be constant (3--5$\times 10^{51}$ erg; e.g. \citet{aretxaga}), in
clear contradiction with our result.
The lifetime of CSNR (0.2--3.8 years) is compatible with the event durations found here. 
But no correlation with the object luminosity is expected, the more luminous objects
simply having higher SNe rates. Again, this is contrary to our results. Finally, \citet{aretxaga} show
that a $-1/2$ slope of the $\sigma(L)$ relation should always be found with this model, which again is not observed. 

\subsubsection{Magnetic blobs above an accretion disk}
\label{magneticblobsaboveanaccretiondisk}

In this model, each event is associated with the discharge of an active magnetic blob
above an accretion disk. In the model proposed by \citet{haardt}, a fraction of the local
accretion power goes into magnetic field structures allowing the
formation of active blobs above the disk. Reconnection of the magnetic
field lines in the corona permits the transfer of the energy into
kinetic energy of fast particles. The energy is stored and released in
the so-called charge and discharge times $t_{\rm c}$ and $t_{\rm d}$ with $t_{\rm d}\ll
t_{\rm c}$. Using the dynamo model of \citet{galeev} for the blob formation, 
\citet{haardt} show that $t_{\rm d}$ scales with the blob size $R_{\rm b}$ which 
itself scales with the total luminosity $L$ of the source. This trend 
is clearly not seen in our
data (Fig.~\ref{Fig7}). Furthermore, the total number of active loops $N_{\rm tot}$, at any
time, does not depend on the luminosity nor on the mass of the object.
The blob rate $N_{\rm tot}/t_{\rm d}$ becomes therefore proportional to $L^{-1}$, also in
clear contradiction with our results that show that the event rate is
not correlated with the luminosity.  Finally, the energy released by a
single blob can be written $E=t_{\rm d}L_{\rm blob}$, where $L_{\rm blob}$ is given	
by Eq.~(7) of \citet{haardt}. The energy $E$ released then
goes with $L^2$, also in contradiction with the results deduced here
in which $E\propto L$.

\subsubsection{Stellar collisions}
\label{stellarcollisions}

\citet{courvoisier2} proposed that the energy radiated in AGN originates in a number of collisions
between stars that orbit the supermassive black hole at very high velocities in a volume of some 100 $R_{\rm S}$.
They computed the rate $\dd n/\dd t$ of head-on stellar
collisions in a spherical shell of width $\dd r$, located at distance $r$ from the central black
hole. The stars are assumed to have mass $M_{\sun}$ and radius $R$ of the Sun. This
reads:
\begin{equation}
\frac{\dd n}{\dd t}=(\rho^24\pi r^2\dd r)v_{\rm K}\pi R^2\;,
\end{equation}
where $\rho$ represents the stellar density and $v_{\rm K}$ the Keplerian 
velocity (\citet{courvoisier2}; \citet{torricelli}). Assuming that the stars
are located in a shell of inner radius $a$ and outer radius $b$ and 
are distributed following a density law $\rho=N_0(r/r_0)^{-\alpha/2}$, with slope $\alpha$ 
where $N_0$ and $r_0$ are constants,
we find for the kinetic energy released by one event:
\begin{equation}\label{meanE}
E=\frac{\int\frac{1}{2}M_{\sun}v_{\rm K}^2\frac{\dd n}{\dd t}}{\int\frac{\dd n}{\dd t}}=\frac{\frac{1}{2}M_{\sun}\int_a ^b\dd r v_{\rm K} ^3(r)r^{2-\alpha}}{\int_a ^b \dd r v_{\rm K}(r)r^{2-\alpha}}\;,
\end{equation}
where $\alpha$, $a$ and $b$ are parameters 
of the stellar cluster. Using $v_{\rm K}=\sqrt{GM_{tot}/r}$ leads to:
\begin{equation}
E=\frac{\frac{1}{2}M_{\sun}G\int_a ^b\dd rM_{tot}^{\frac{3}{2}}r^{\frac{1}{2}-\alpha}}{\int_a ^b \dd rM_{tot}^{\frac{1}{2}}r^{\frac{3}{2}-\alpha}}\;.
\end{equation}
Neglecting the cluster's mass with respect to the black hole mass,
i.e. making the assumption that $M_{\rm tot}(r)\sim M_{\rm BH}$, we finally
have:
\begin{equation}
E=\frac{1}{2}M_{\sun}GM_{\rm BH}f(\alpha,a,b)\;,
\end{equation}
where $f$ is a function of the cluster parameters only, independent of $M_{\rm BH}$.

We need now to relate the average luminosity to the black hole mass.
This relation comes from $\overline{L}=\int 1/2 M_{\sun} v_{\rm K}^2 dn/dt= (M_{\sun}R_{\sun}^2\pi^2G^{3/2}N_0r_0^{\alpha})M_{\rm BH}^{3/2}f(\alpha,a,b)$.
As $E\propto M_{\rm BH}$ and $\overline{L}\propto M_{\rm BH}^{3/2}$, we
finally have:
\begin{equation}
E\propto\overline{L}^{\frac{2}{3}}\;.
\end{equation}
This relation is a relatively good approximation of the trends seen in Fig.~\ref{Fig11}, 
although not completely satisfactory.

In this model, the variability time scale $2\mu_{\lambda}$ is expected
to be related to the time needed to the expanding sphere to become
optically thin. This point is discussed in \citet{courvoisier3} who found that for clumps of 
about one Solar mass, the expansion time is about $2\times 10^6$ seconds. This enters in the
range of time scales found here.  
The collision rate should be going with $M_{\rm BH}^{1/2}$, which implies
$N\propto \overline{L}^{1/3}$, which is not seen in our data.

\subsubsection{Other models}
\label{othermodels}

For the sake of completeness, we note that gravitational micro-lensing
models, in which populations of planetary mass compact bodies randomly
cross the line of sight of an observed AGN, can be invoked
to explain the long term variations over several
years. But in the case of low-redshift Seyfert galaxies, which forms
the majority of our sample, the probabilities of micro-lensing are not
significant \citep{hawkins}.

Finally, a model of accretion disk instabilities has been suggested
\citep{kawaguchi} to explain the optical variability of AGN. The SOC
state model \citep{mineshige} is producing power density spectra in
good agreement with the observations but, since it is not an
event-based model, it is difficult to use the measurements discussed
here to constrain it.

\section{Conclusion}
\label{summaryandconclusion}

We showed the existence of a maximum variability time scale in the ultraviolet light curves of
15 Type 1 AGN, in the range 1300--3000 \AA. We found variability time scales in the range 0.02--1.00 year.

In the framework of the discrete-event model, we showed that these 
time scales can be related to the event duration in a simple
manner. 
A weak dependence of the event duration with the object luminosity at 1300 \AA\ is found. The event duration is not a function of the wavelength 
in the range 1300 to 3000 \AA. 

The event energy per object varies from $10^{48}$ to $10^{52}$ erg with 
a corresponding event rate comprised between 2 and 270 events per year, assuming the 
presence of a constant component in the light curves.

Our results do not depend on the constant component $C_{\lambda}$. 
While we can only provide lower and upper bounds on $C_{\lambda}$, its choice does not
change the conclusions.

The event energy is strongly correlated with the object luminosity. We show that the combined
relations of the event energy $E\propto\overline{L_{1300}}^{1.02}$, and event 
duration $2\mu_{1300}\propto\overline{L_{1300}}^{0.21}$
with the object luminosity,
lead to the trend seen in the variability-luminosity relationship in the rest 
frame, i.e. that both variables are correlated with a slope of about 0.08.
We thus established that the event parameter 
which drives the $\sigma(L)$ dependence is the event energy, and not its duration, nor its rate.

These results allow us to constrain the physical nature of the events. We show that neither 
the starburst model nor the magnetic blob model can satisfy these requirements. On the other 
hand, stellar collision models in which the average properties of the collisions depend on 
the mass of the central black hole may be favored, although the model will need to be improved as the
result we found (for instance, the lack of correlation between the event rate and the luminosity)
does not match the predictions.

\begin{acknowledgements}
SP acknowledges a grant from the Swiss National Science Foundation.
\end{acknowledgements}

\bibliographystyle{aa}  
\bibliography{biblio} 

\appendix 
\section{Structure functions 1300--3000 \AA}
(Online material)
\end{document}